\documentclass[preprint2]{aastex}
\usepackage{emulateapj5}
\usepackage{onecolfloat5}

\newcommand{\beq}   {\begin{equation}}
\newcommand{\eeq}   {\end{equation}}

\newcommand{\kms}   {km~s$^{-1}$}
\newcommand{\water}   {H$_2$O}
\newcommand{\aam}   {\altaffilmark}
\newcommand{\aat}   {\altaffiltext}

\shortauthors{Vlemmings \& Diamond}
\shorttitle{Properties of the W43A H$_2$O maser jet}
\slugcomment{Astrophysical Journal Letters accepted 07/14/06}

\begin{document}
\twocolumn[%%% Begin front material
\title{Intrinsic properties of the magnetically collimated H$_2$O maser jet of W43A}
\author{ W. H. T. Vlemmings\aam{1}, P. J. Diamond\aam{1}}

\begin{abstract}
 Water maser polarization observations in the precessing jet of W43A
 have revealed that it is magnetically collimated. Here we present a
 detailed description of the physical properties of the water maser
 environment in the jet. We discuss the maser saturation level and
 beaming angle as well as the intrinsic temperatures and
 densities. Additionally, we show that the polarization angle of the
 strongest red-shifted maser feature undergoes a fast rotation of
 $90^\circ$ across the maser. Along with the variation of linear
 polarization fraction, this strongly supports the current theoretical
 description of maser linear polarization.
\end{abstract}
\keywords{Stars:individual (W43A)---magnetic fields---polarization---masers}
]%%% End front material

\aat{1}{Jodrell Bank Observatory, University of Manchester, Macclesfield, Cheshire SK11 9DL, U.K.; wouter@jb.man.ac.uk}

\section{Introduction}
W43A is an evolved star at a distance of 2.6~kpc \citep{Diamond85} and 
is surrounded by a thick circumstellar envelope (CSE) that exhibits
OH, \water\ and SiO masers \citep[][~and references
  therein]{Imai05}. Unlike the shell-like structure of the 22~GHz
\water\ masers typically found in the envelopes of evolved stars, the
\water\ masers of W43A occur in two clusters at 1000 AU from the star
near the opposing tips of a collimated jet. The jet, with a velocity
of 145 \kms, has an inclination of 39$^\circ$ with respect to the sky
plane, a position angle of 65$^\circ$ and shows a 5$^\circ$ precession
with a period of 55 years. The inferred dynamical age of the jet is only approximately 50 years \citep{Imai02}.  W43A is interpreted as
belonging to a class of objects undergoing a rapid transition from an
evolved star into a planetary nebula (PNe). Owing to their short
expected lifetime of less than 1000 years, only 4 sources of this
class have been identified to date \citep{Imai02, Imai04, Likkel92, Morris03,
  Boboltz05}.  \water\ masers at 22 GHz are excited in gas with
temperatures of $T\approx 400$K and hydrogen number densities of $n =
10^8-10^{10}$ cm$^{-3}$ \citep{Elitzurbook}. These conditions are
typically found close to the star. The \water\ masers in the
collimated jet of W43A, however, likely arise when the jet has swept up
enough material previously expelled from the star so that conditions
at the tip of the jet have become favorable for \water\ masers to
occur. Alternatively, they occur in a shock between the collimated
jet and dense material in the outer CSE, similar to the
\water\ masers found in star-forming regions.

Observations of linear and circular polarization of the different
maser species in CSEs are uniquely suited to study the strength and
structure of magnetic fields. Close to the central star, at radii of
5--10 AU, SiO masers indicate ordered fields of the order of several
Gauss \citep[e.g.][]{Kemball97, Herpin06}. At the outer edge of the
CSE, the polarization measurements of OH masers reveal milliGauss
magnetic fields and indicate weak alignment with CSE structure
\citep[e.g][]{Etoka04}. Recently, the Zeeman splitting giving rise to
the circular polarization of \water\ masers was measured for a sample
of evolved stars \citep{Vlemmings02, Vlemmings05b}. It was shown that
at distances of several tens to hundreds of AU, CSEs harbor large
scale magnetic fields with typical field strengths between a few
hundred milliGauss up to a few Gauss. While the origin of the magnetic
field is still unclear, theoretical models have shown that a dynamo
between the slowly rotating stellar outer layers and the faster
rotating core can produce the observed magnetic fields
\citep{Blackman01}. However, this likely requires an additional source
of angular momentum to maintain the magnetic field, which could be
provided by the presence of a binary companion or heavy planet
\citep{Blackman04}.

Magnetic fields around evolved stars are thought to be
one of the main factors in shaping the CSEs and producing the
asymmetries during evolution of a spherically symmetric star into
the often asymmetric PN. Theoretical models show that magnetic fields
could be the collimating agents of the bi-polar jets in young
proto-planetary nebulae such as W43A \citep[e.g][]{Garcia05}. In a
recent paper, we have shown, using \water\ maser linear and circular
polarization observations, that the magnetic field is indeed the
collimating agent of the jet of W43A \citep[][~hereafter paper
  I]{Vlemmings06b}. Here we present a more detailed analysis of the
observations presented in paper I. We discuss the physical properties
of the maser region. We also show how the observed linear polarization
characteristics strongly support the current maser theory.

\begin{figure*}[ht!]
\epsscale{2.0} 
%\epsscale{1.0} 
\plottwo{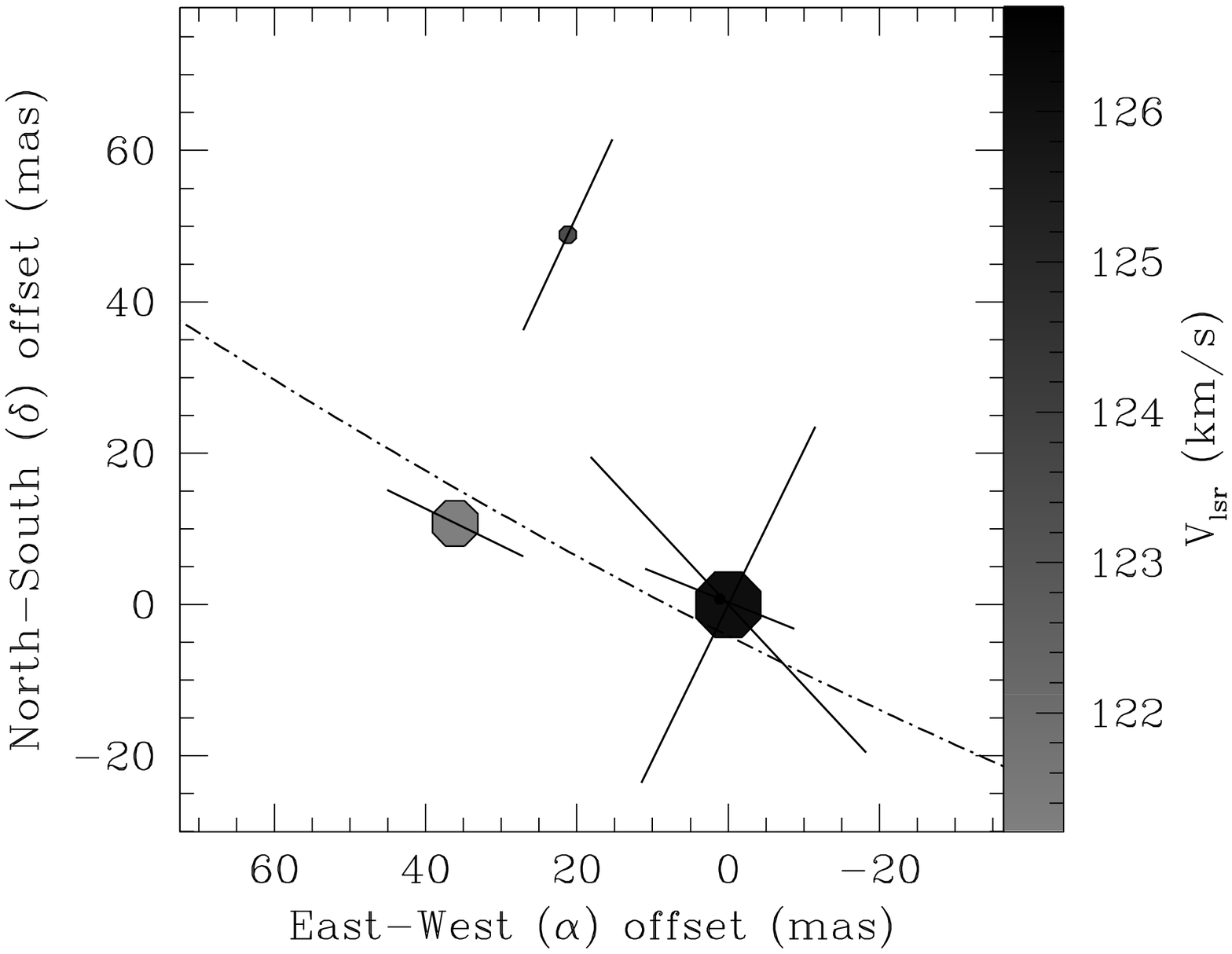}{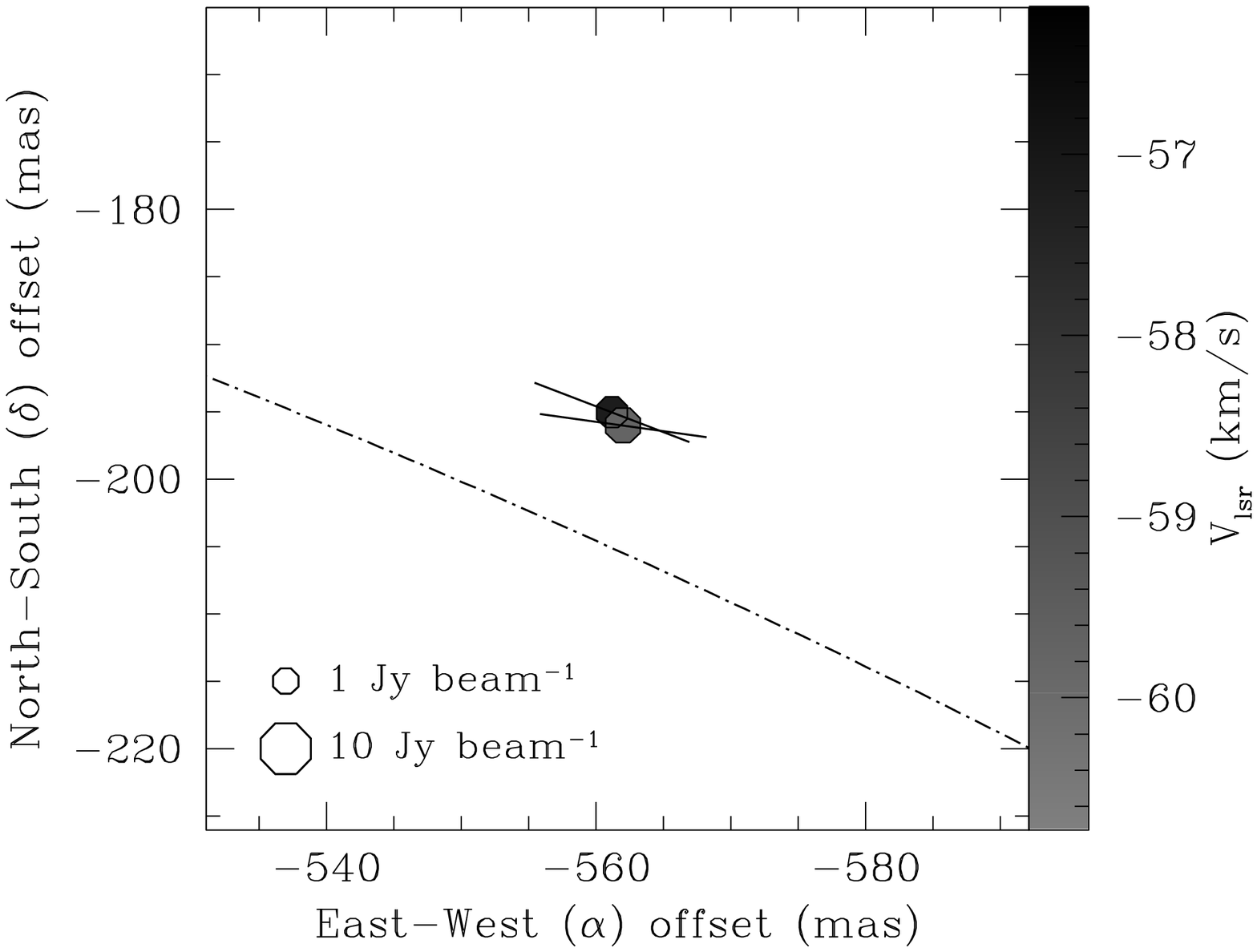}
\caption[maps]{The \water\ maser features for which polarization was
  detected in the red- (left) and blue-shifted (right) tips of the jet
  of W43A. The symbols are scaled logarithmically according to peak
  flux and the grey-scale indicates their velocity. The linear
  polarization vectors are scaled logarithmically as a function
    of polarization percentage. The dashed-dotted line indicates the
  precessing jet model.}
\label{Fig:maps}
\end{figure*}

\section{Observations}

The observations of W43A were performed with the NRAO\footnote{The
  National Radio Astronomy Observatory (NRAO) is a facility of the
  National Science Foundation operated under cooperative agreement by
  Associated Universities, Inc.} VLBA on December 8 2004 at the
frequency of the $6_{16} - 5_{23}$ rotational transition of H$_2$O:
22.235080 GHz.  We used 4 baseband filters of 1~MHz width, which were
pair-wise overlapped to get a velocity coverage of
$\approx25$~\kms\ around both the blue-shifted ($V_{\rm
  lsr}=-56$~\kms) and red-shifted ($V_{\rm lsr}=124$~\kms) tip of the
\water\ maser jet. Similar to the observations of \water\ masers in
star-forming regions described in \citet[][~hereafter
  V06]{Vlemmings06}, the data were correlated multiple times with a
correlator averaging time of 8~sec. The initial correlation was
performed with modest spectral resolution (128 channels; $7.8$~kHz$ =
0.1$~\kms), which enabled us to generate all 4 polarization
combinations (RR, LL, RL and LR). Two additional correlator runs were
performed with high spectral resolution (512 channels; $1.95$~kHz$ =
0.027$~\kms), which therefore only contained the two polarization
combinations RR and LL, to be able to detect the signature of the
H$_2$O Zeeman splitting across the entire velocity range. The
observations on W43A were interspersed with 15 minute observations of
the polarization calibrator J1743-0350. Including scans on the phase
calibrators (3C345 and 3C454.3), the total observation time was 8
hours. The data analysis path is described in detail in
\citet{Vlemmings02} and follows the method of
\citet{Kemball95}. Fringe fitting and self-calibration were performed
on a strong ($\sim$35~Jy beam$^{-1}$) maser feature (at $V_{\rm
  lsr}=-15.72$~\kms). Image cubes were created of Stokes I,Q and U
from the intermediate spectral resolution data and of Stokes I and V
with high spectral resolution. Due to the low declination of W43A
($\alpha{\rm (J2000)}=18h47m41.166s$ and $\delta{\rm
  (J2000)}=-01{^\circ}45{\arcmin}11.7{\arcsec}$), the beam width is
$\approx~0.6 \times 1.3$~mas. In the high spectral resolution total
intensity and circular polarization channel maps, the noise is
  $\approx$14~mJy beam$^{-1}$. In the lower spectral resolution Stokes
  Q and U maps the rms noise is $\approx~10$~mJy beam$^{-1}$. We
estimate our polarization angles to contain a possible systematic
error of $\sim$8$^\circ$ due to the error in the polarization
  angle of the calibrator J1743-0350.

%\begin{landscape}
\begin{deluxetable}{cccccccccc}
\tablecolumns{10}
\tablewidth{0pc} 
\tablecaption{W43A maser features\label{Table:masers}}
\tablehead{ 
\colhead{feature} & \colhead{$\Delta\alpha$} & \colhead{$\Delta\delta$} &
\colhead{peak intensity} & \colhead{$\Delta v$} & \colhead{$V_{\rm lsr}$} & \colhead{$P_L$} & \colhead{$\chi$} & \colhead{$P_V$} & \colhead{$B\cos\theta$}\\
\colhead{} & \colhead{(mas)} & \colhead{(mas)} &
\colhead{(Jy beam$^{-1}$)} & \colhead{(\kms)} & \colhead{(\kms)} & \colhead{($\%$)} & \colhead{($^\circ$)} & \colhead{($\%$)} & \colhead{(mG)}
}
\startdata
 1$^a$ & 0.0 & 0.0 & 38.57 & 0.54 & 126.11 & $2.29 \pm 0.40$ & $54 \pm 3$ & - & $<36$\\
   & & & & & & $2.25 \pm 0.42$ & $-29 \pm 1$& & \\
 2 & 1.1 & 0.7 & 4.89 & 0.68 & 126.69 & $0.62 \pm 0.23$ & $68 \pm 10$ & - & $<278$\\
 3 & 21.2 & 48.9 & 6.45 & 0.79 & 123.42 & $0.82 \pm 0.23$ & $-25 \pm 12$ & - & $<306$\\
 4 & 36.1 & 10.7 & 18.66 & 0.75 & 121.21 & $0.59 \pm 0.17$ & $64 \pm 4$ & - & $<101$\\
 5 & -561.2 & -195.0 & 10.85 & 0.88 & -57.19 & $0.64 \pm 0.19$ & $69 \pm 8$ & $0.33 \pm 0.09$ & $85 \pm 33$ \\
 6 & -562.0 & -196.0 & 12.45 & 0.66 & -59.73 & $0.65 \pm 0.06$ & $82 \pm 6$ & - & $<93$\\
\enddata 
\tablecomments{$^a$ polarization angle gradient across maser feature (see text)}
\end{deluxetable} 
%\end{landscape}

\section{Results}

We detected polarization in several 22~GHz \water\ maser features in
both the red-shifted and blue-shifted tips of the jet of W43A. The
complete map of detected maser features is shown in Figure 1 of paper
I. Here we show, in Fig.~\ref{Fig:maps}, the maser features for which
linear polarization was detected. In Table~\ref{Table:masers} we list
the position off-sets with respect to the reference feature
($\Delta\alpha$ and $\Delta\delta$), peak flux, full-width half
maximum line-width ($\Delta v$) and LSR velocity ($V_{\rm lsr}$) of
these maser features. The table also contains the fractional linear
polarization ($P_L$), polarization angle ($\chi$), fractional circular
polarization ($P_V$) and magnetic field strength ($B$) determined at
angle $\theta$ from the line-of-sight. The polarization
  properties and corresponding rms errors are determined from a flux
  weighted average over $\Delta v$ for each maser line. We only
detected circular polarization, at a level of $3.7\sigma$, in one of
the blue-shifted maser features (feature $5$), and its polarization
spectrum is shown in Figure 2 of paper I. The magnetic field strength
and $3\sigma$ field strength upper limits were determined as described
in V06. This analysis resulted in an estimate of $B\cos\theta=85 \pm
33$~mG for the feature where circular polarization was detected.

\begin{figure*}[ht!]
\epsscale{1.5} 
%\epsscale{1.0} 
\plottwo{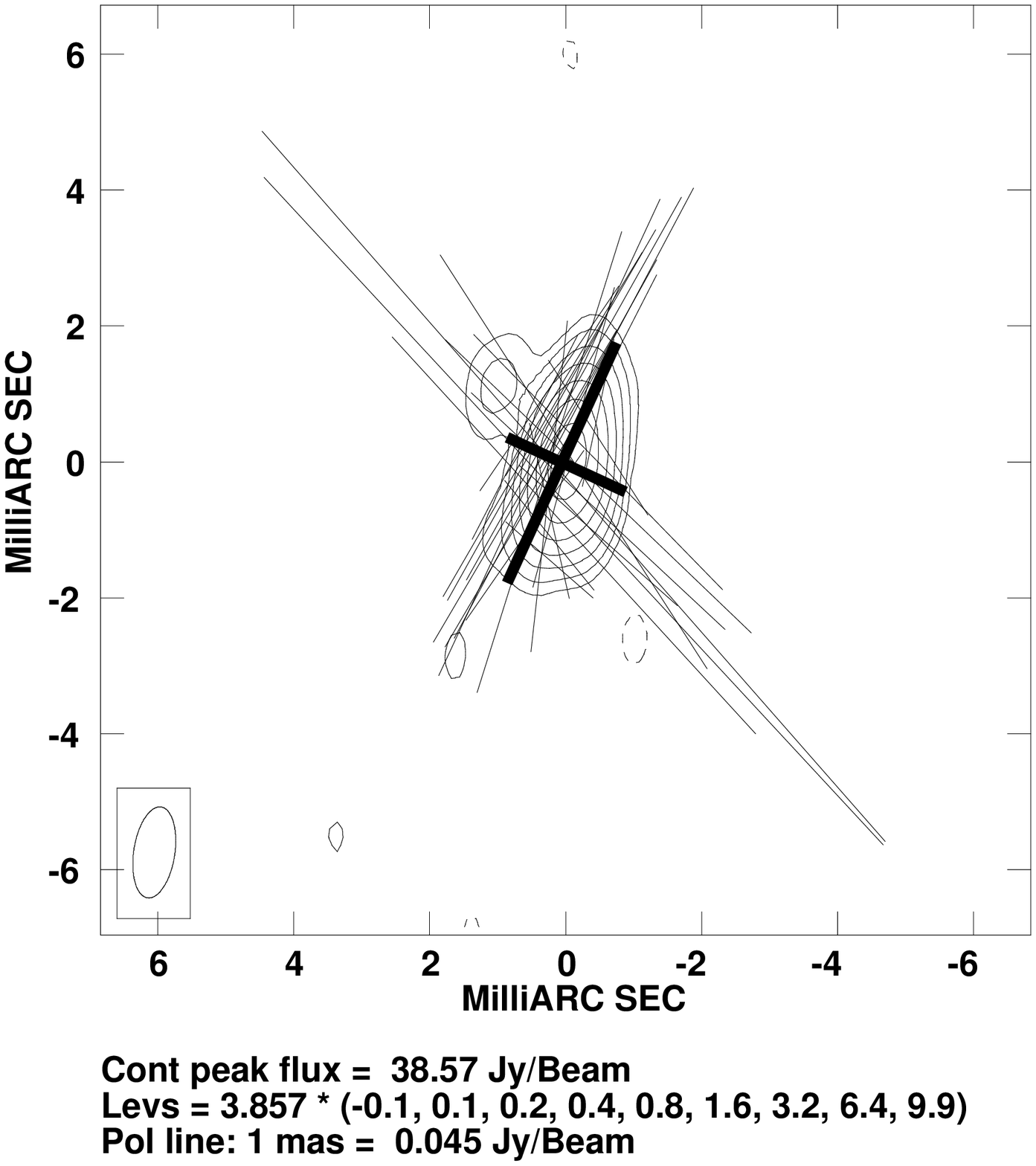}{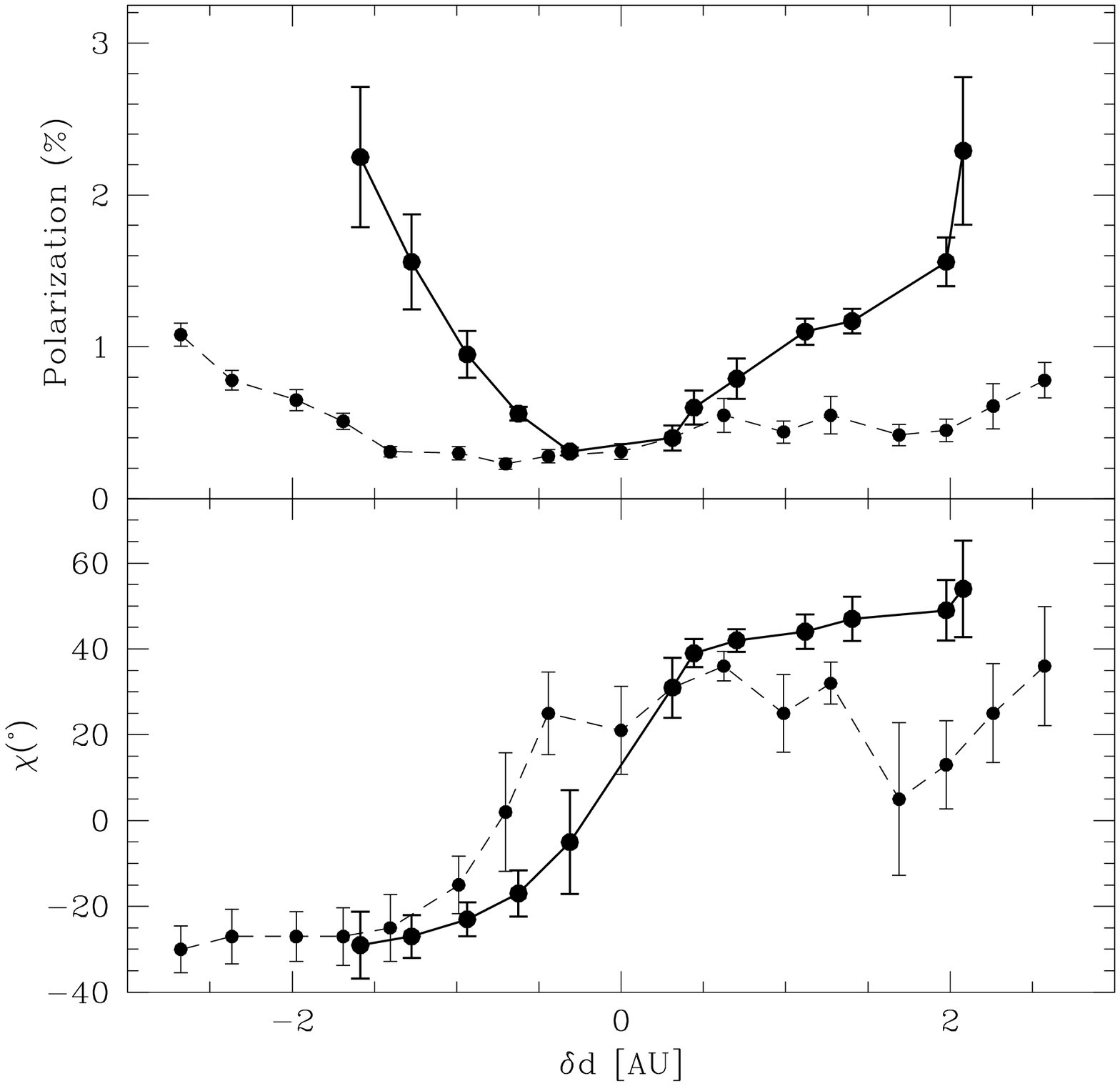}
\caption[lpol]{(left) Total intensity map of the strongest maser
  feature in the red-shifted tip of the jet of W43A at $V_{\rm
    lsr}=126.11$~\kms. The vectors indicate the linear polarization,
  the thick solid lines indicate the slices across the feature for
  which linear polarization angle and fractional linear polarization
  are shown at the right. (right) Fractional linear polarization (top)
  and polarization angle $\chi$ (bottom) versus angular offset for
  slices across the strongest maser feature. The offset $\delta$d is
  determined with respect to the position of the peak maser emission
  and assuming a distance to W43A of 2.6~kpc. $\delta$d increases
  towards the South. The thin dashed lines indicate the slice
  perpendicular to the jet direction at PA=$-27^\circ$ and the thick
  solid line indicates the slice along the jet direction at
  PA=$63^\circ$.}
\label{Fig:lpol}
\end{figure*}

We have detected linear polarization in the six brightest maser
features.  As discussed in paper I, the weighted mean linear polarization
fraction of these masers is $0.66 \pm 0.07 \%$ and the $3\sigma$ upper
limits in the masers where no linear polarization was detected range
upwards from $0.68\%$. In the brightest of the maser features
(feature $1$), we observe a 90$^\circ$ flip of polarization angle
across the maser, which is shown in Fig.~\ref{Fig:lpol}. This is not
observed for any of the other maser features. The fractional
polarization of feature $1$ also varies across the maser and the
values in Table.~\ref{Table:masers} represent the values at the
location of the maximum and minimum polarization angle. The flux
weighted average of the linear polarization fraction of the maser
feature is $0.99 \pm 0.49 \%$.

\section{Discussion}

\subsection{Intrinsic properties of the maser regions}

As described in V06, the models used to determine the magnetic field
from the maser total intensity and polarization spectra also yield the
intrinsic maser thermal width $\Delta v_{\rm th} \approx
0.5(T/100)^{1/2}$, where $T$ is the temperature in the maser
region. Additionally, the models produce the maser emerging brightness
temperature $T_b\Delta\Omega$, where $T_b$ is the brightness
temperature and $\Delta\Omega$ the unknown beaming solid angle. Circular polarization measurements are necessary to optimally constrain $\Delta v_{\rm th}$ and $T_b\Delta\Omega$, and we have been able to
determine $\Delta v_{\rm th}$ and $T_b\Delta\Omega$ for one of the
maser features. We find that the intrinsic thermal width of feature
$5$, in the blue-shifted tip of the jet, is $\Delta v_{\rm th}=1.1 \pm
0.3$~\kms. This indicates an intrinsic temperature in the maser jet of
$T \approx 500$~K. For the emerging brightness temperature we find
$T_b\Delta\Omega \approx 8\times10^9$~K~sr.

The masers in the collimated jet of W43A are found at $\sim 1000$~AU
from the central star, much further out than is typical for the
\water\ masers in the CSE of evolved stars. They likely arise when
the jet has swept up enough material to create suitable conditions for
maser excitation. Alternatively, the masers occur in a shocked region
between the jet and dense material in the outer CSE, similar to the
shocked masers found in star-forming regions. If the masers occur in a
shock, the temperature in the maser region indicates that it is likely
a dissociative J-shock \citep{Kaufman96}. Using the \water\ maser
models developed for J-shock masers in star-forming regions
\citep{Elitzur89}, we then find that the pre-shock hydrogen density
would be $n \sim 3\times 10^6$~cm$^{-3}$. This is an order of
magnitude higher than the typical hydrogen density at $\sim 1000$~AU
from the star ($n\sim10^5$~cm$^{-3}$). We thus conclude that the
\water\ masers are likely excited in material swept up in the jet
instead of in a J-shock.

We can compare the emerging brightness temperature determined from the
model, with the maser brightness temperature determined from the
observations. We find that feature $5$ is unresolved. Assuming a size
of $\sim 0.4$~mas, which corresponds to $\approx 1$~AU, we find a
brightness temperature lower limit $T_b\gtrsim
1.5\times10^{11}$~K. Thus, the upper limit on the beaming solid angle
$\Delta\Omega\lesssim 5\times10^{-2}$~sr, which is similar to the
values found in star-forming regions (V06). Most of the other maser
features are also unresolved, with only the brightest feature (feature
$1$) being marginally resolved. Thus, the W43A \water\ masers have
brightness temperatures of $T_b\approx 10^{11} - 10^{12}$~K.
Additionally, we can compare the emerging brightness temperature with
the maser brightness temperature $T_S$ at the onset of
saturation. Using \citet{Reid88}, we find
$T_S\Delta\Omega=3.4\times10^9$~K~sr. This indicates, for
$\Delta\Omega\approx 5\times10^{-2}$~sr, that the masers of W43A are
mostly saturated.

\subsection{Linear Polarization}

Maser theory has shown that the percentage of linear polarization
$P_L$ of \water\ masers depends on the degree of saturation and the
angle $\theta$ between the maser propagation direction and the
magnetic field \citep[e.g.][]{Deguchi90}. Additionally, when the
Zeeman frequency shift $g\Omega$ is much greater than the rate for
stimulated emission $R$, the polarization vectors are either
perpendicular or parallel to the magnetic field lines, depending on
$\theta$. For the typical emerging brightness temperature in the
\water\ maser jet of W43A, $T\Delta\Omega\approx 10^{10}$~K~sr,
$R\approx 1$~s$^{-1}$, while for a magnetic field of $\approx 100$~mG,
the strongest 22~GHz hyperfine transitions have $g\Omega\sim
1000$~s$^{-1}$. Thus $g\Omega >> R$ is easily satisfied. As a result,
when $\theta>\theta_{\rm crit}\approx 55^\circ$ the polarization
vectors are perpendicular to the magnetic field, and when
$\theta<\theta_{\rm crit}$ they are parallel \citep{Goldreich73}.  The
strongest linear polarization is found when $\theta>\theta_{\rm
  crit}$, thus it was concluded in paper I that most of the linear
polarization vectors were perpendicular to the magnetic field,
indicating a toroidal magnetic field configuration. If alternatively
they are parallel to the magnetic field we are actually probing the
poloidal component.

When $\theta$ is close to $\theta_{\rm crit}$, the linear polarization
vectors can flip $90^\circ$ on very small scales. According to maser
theory, a minimum in $P_L$ occurs when $\theta=\theta_{\rm
  crit}$. This can be seen in figure A.1 of V06, which gives $P_L$ as a
function of $\theta$ and saturation level. Here, we observe a
$90^\circ$ flip across the brightest of the maser features in the jet
of W43A, as shown in Fig.~\ref{Fig:lpol}. Similar flips were observed
in circumstellar SiO masers \citep{Kemball97} and \water\ masers in
star-forming regions (V06). However, here we have for the first time
been able to detect the expected minimum in $P_L$ when the flip
occurs. The minimum is most pronounced in the slice across the maser
along the PA of the jet. The linear polarization characteristics are
dominated by the condition along the maser path where the strongest
amplification occurs \citep{Vlemmings06c}. Thus, we conclude that the
strongest amplification in the masing region giving rise to feature
$1$ occurs at the location along the line-of-sight where $\theta$ is
close to $\theta_{\rm crit}$. The position angle changes from $\chi=54^\circ$, which is
perpendicular to the magnetic field, to $\chi=-29^\circ$, which is parallel
to the magnetic field. Most of the other maser
features have polarization angles that are perpendicular to the
magnetic field except possibly feature $3$ with $\chi=-25^\circ$, possibly
because this maser feature is located at the edge of the jet, where a
projected toroidal magnetic field will lie along the jet.

\subsection{Magnetic Field}

After correcting for the angle between the line-of-sight and the
magnetic field, and taking into account the saturation level, we find
a toroidal magnetic field of $B\approx200$~mG in the \water\ maser jet
of W43A (paper I). The magnetic field is increased in the density-enhanced \water\ maser jet due to partial coupling to the gas. The
magnetic field outside the jet is found to be $B\approx0.9-2.6$~mG
if the masers exist in swept up material at a density of
$n=10^8-10^{10}$~cm$^{-3}$. If the masers are excited in a
dissociative shock, the magnetic field outside the jet is
$B\approx0.07$~mG, however, as discussed above, this is unlikely to be
the case. In paper I it was shown that if we are probing the toroidal
field, with $B_\phi\propto r^{-1}$, this implies an average magnetic
field of $B\approx20$~G on the surface of the star. This is in
excellent agreement with the magnetic field values determined by
extrapolating the field strengths found using SiO, \water\ and OH
maser observations throughout the CSEs of a large number of evolved
stars. If, instead of the toroidal field, we are probing the poloidal
field (with its radial component $B_r\propto r^{-2}$), the surface
magnetic field would be a factor of $\sim 10^3$ larger. As this is
inconsistent with the other observations, we conclude that the
\water\ maser polarization vectors are indeed tracing the toroidal
magnetic field.

\section{Conclusions}

Using polarization observations of the \water\ masers in the tips of
the jet of W43A we have shown that the masers most likely occur in
swept up material, and are not shock excited. The masers are saturated
with a typical beaming solid angle of $\Delta\Omega\approx
5\times10^{-2}$~sr and occur at temperatures of $T\approx 500$~K. The
polarization vectors trace the toroidal magnetic field, which has a
density enhanced field strength of $B\approx200$~mG. This implies, as
shown in paper I, a magnetic field strength that is sufficiently strong
to produce a magnetically collimated jet.  Additionally, a $90^\circ$
flip of the polarization angle and a corresponding minimum in
fractional linear polarization is observed across the brightest maser
feature. This strongly supports the current maser polarization theory.

\begin{acknowledgements}
This research was supported by a Marie Curie Intra-European fellowship
within the 6th European Community Framework Program under contract
number MEIF-CT-2005-010393.
\end{acknowledgements}

%\bibliographystyle{aa}

%\bibliography{wvrefs}

%\thispagestyle{headings}

\end{document}